\documentclass[aps,eqsecnum,preprint,preprintnumbers,12pt,amsfonts]{revtex4}

\usepackage{epsfig,psfrag}
\usepackage{latexsym}
\usepackage{graphicx}
\usepackage{axodraw}

\def\beq{\begin{eqnarray}}
\def\eeq{\end{eqnarray}}
\def\bea{\begin{eqnarray}}
\def\eea{\end{eqnarray}}
\def\det{{\rm det}}
\newcommand\eqn[1]{(\ref{#1})}      
\newcommand{\nn}{\nonumber}
\newcommand{\reals}{\mbox{${\rm I\!R }$}}

\newcommand{\ep}{\epsilon}

\newcommand{\intl}{\int\limits}

\usepackage{color}

\def\firstorder{\begin{picture}(10,10)(10,10)
\CArc(15,15)(10,0,180)
\CArc(15,15)(10,180,360)
\Vertex(5,15)2
\end{picture}}

\def\secondorder{\begin{picture}(10,10)(10,10)
\CArc(15,15)(10,0,180)
\CArc(15,15)(10,180,360)
\Vertex(5,15)2
\Vertex(25,15)2
\end{picture}}

\def\thirdorder{\begin{picture}(10,10)(10,10)
\CArc(15,15)(10,0,180)
\CArc(15,15)(10,180,360)
\Vertex(15,25)2
\Vertex(24,12)2
\Vertex(6,12)2
\end{picture}}

\begin{document}

\title{Functional determinants for radial operators}
\author{Gerald V. Dunne}\email{dunne@phys.uconn.edu}
\affiliation{Department of Physics,
University of Connecticut, Storrs, CT 06269}
\author{Klaus Kirsten}\email{klaus_kirsten@baylor.edu}
\affiliation{Department of Mathematics, Baylor University,
Waco, TX 76798}


\begin{abstract}
We derive simple new expressions, in various dimensions, for the
functional determinant of a radially separable partial
differential operator, thereby generalizing the one-dimensional
result of Gel'fand and Yaglom to higher dimensions. We use the
zeta function formalism, and the results agree with what one would
obtain using the angular momentum cutoff method based on radial
WKB. The final expression is numerically equal to an alternative
expression derived in a Feynman diagrammatic approach, but is
considerably simpler.
\end{abstract}

\maketitle

\section{Introduction and Results}
\label{sec-intro}

Determinants of differential operators occur naturally in many
applications in mathematical and theoretical physics, and also
have inherent mathematical interest since they encode certain
spectral properties of differential operators. Physically, such
determinants arise in semiclassical and one-loop approximations in
quantum mechanics and quantum field theory
\cite{schwinger,salam,coleman}. Determinants of free Laplacians
and free Dirac operators have been extensively studied
\cite{ray,hawking,dhoker,sarnak,stuart1,elizalde,kirstenbook},
but much less is known about operators involving an arbitrary
potential function. For ordinary ({\it i.e.}, one dimensional)
differential operators, a general theory has been developed for
determinants of such operators
\cite{gy,levit,forman,kirstenbook,kirsten,kleinert}. In this paper
we extend these results to a broad class of separable {\it
partial} differential operators. The result for four dimensions
was found previously in \cite{dunnemin} using radial WKB and an
angular momentum cut-off regularization and renormalization
\cite{wkb}. Here we re-derive this result using the zeta function
approach to determinants \cite{ray,hawking,elizalde,kirstenbook},
and generalize to other dimensions. We are motivated by
applications in quantum field theory, so we concentrate on  the
dimensions $d=2, 3, 4$, but the mathematical extension to higher
dimensions is immediate. We also compare with another expression
for the four dimensional determinant \cite{baacke}, derived using
a Feynman diagrammatic approach.

Consider the radially separable partial differential operators
\bea
{\mathcal M}&=&-\Delta+V(r)
\label{op} \\
{\mathcal M}^{\rm free}&=&-\Delta\quad , \label{free-op} \eea
where $\Delta$ is the Laplace operator in $\reals^d$, and $V(r)$
is a radial potential vanishing at infinity as $r^{-2-\epsilon}$
for $d=2$ and $d=3$, and as $r^{-4-\epsilon}$ for $d=4$. For
$d=1$, with Dirichlet boundary conditions on the interval $[0,
\infty)$, the results of Gel'fand and Yaglom \cite{gy} lead to the
following simple expression for the determinant ratio: \bea
\frac{\det \left[{\mathcal M}+m^2\right]}{\det \left[{\mathcal
M}^{\rm free}+m^2\right]} =\frac{\psi(\infty)}{\psi^{\rm
free}(\infty)}\quad , \label{gy-result} \eea where
$\left[{\mathcal M}+m^2\right]\psi=0$, with {\it initial value}
boundary conditions: $\psi(0)=0$ and $\psi^\prime(0)=1$. The
function $\psi^{\rm free}$ is defined similarly in terms of the
free operator: $[{\mathcal M}^{\rm free}+m^2]$. The squared mass,
$m^2$, is important for physical applications, and plays the
mathematical role of a spectral parameter.  The result
\eqn{gy-result} is geometrically interesting, in addition to being
computationally simple, as it means that the determinant is
determined simply by the {\it boundary values} of the solutions of
$\left[{\mathcal M}+m^2\right]\psi=0$, and no detailed information
is needed concerning the actual spectrum of eigenvalues.

Now consider dimensions greater than one (as mentioned, we are
most interested in $d=2, 3, 4$; but the extension to higher
dimensions is immediate). Since the potential is radial, $V=V(r)$,
we can express the eigenfunctions of ${\mathcal M}$ as linear
combinations of basis functions of the form: \bea \Psi(r,
\vec{\theta})=\frac{1}{ r^{(d-1)/2}} \, \psi_{(l)}(r)\,
Y_{(l)}(\vec{\theta})\quad , \label{separation} \eea where
$Y_{(l)}(\vec{\theta})$ is a hyperspherical harmonic
\cite{gradsteyn}, labeled in part by a non-negative integer $l$,
and the radial function $\psi_{(l)}(r)$ is an eigenfunction of the
Schr\"odinger-like radial operator \bea {\mathcal M}_{(l)}\equiv
-\frac{d^2}{dr^2}+\frac{\left(l+\frac{d-3}{2}\right)\left(l+\frac{d-1}{2}\right)}{r^2}+V(r)\quad
. \label{radial-op} \eea ${\mathcal M}_{(l)}^{\rm free}$ is
defined similarly, with the potential omitted: $V=0$. In dimension
$d\geq 2$, the radial eigenfunctions $\psi_{(l)}$ have degeneracy
given by \cite{gradsteyn} \bea {\rm deg}(l ; d)\equiv
\frac{(2l+d-2)(l+d-3)!}{l!(d-2)!} \quad . \label{deg} \eea
Formally, for the separable operators in \eqn{op}--\eqn{free-op},
the logarithm of the determinant  ratio can be written as a sum
over $l$ (weighted with the degeneracy factor) of the logarithm of
one-dimensional determinant ratios, \bea \ln\left(\frac{\det
\left[{\mathcal M}+m^2\right]}{\det \left[{\mathcal M}^{\rm
free}+m^2\right]}\right) =\sum_{l=0}^\infty {\rm deg}(l; d)
\ln\left(\frac{\det \left[{\mathcal M}_{(l)}+m^2\right]}{\det
\left[{\mathcal M}^{\rm free}_{(l)}+m^2\right]}\right) \quad .
\label{formal-sum} \eea Each term in the sum can be computed using
the Gel'fand-Yaglom result \eqn{gy-result}. However, the $l$ sum
in \eqn{formal-sum} is divergent, as noted by Forman \cite{forman}
for the free Laplace operator in a two-dimensional disc. In this
paper we show how to define a finite and renormalized determinant
ratio for the radially separable partial differential operators
\eqn{op}--\eqn{free-op}. Specifically, we derive the following
simple expressions, which generalize \eqn{gy-result} to higher
dimensions:
\bea
\ln\left(\frac{\det [{\mathcal M}+m^2]}{\det [{\mathcal M}^{\rm free}+m^2]}\right)\Bigg |_{d=2}
&=&\ln \left(\frac{\psi_{(0)}(\infty)}{\psi^{\rm free}_{(0)}(\infty)}\right)+
\sum_{l=1}^\infty 2 \left\{\ln \left(\frac{\psi_{(l)}(\infty)}{\psi^{\rm free}_{(l)}(\infty)}\right)-\frac{\int_0^\infty dr\, r V(r)}{2l}\right\} \nn\\
&&+\int_0^\infty dr\, r\, V \left[\ln\left(\frac{\mu r}{2}\right)+\gamma \right]
\label{2d-result}\\ \nn\\
\ln\left(\frac{\det [{\mathcal M}+m^2]}{\det [{\mathcal M}^{\rm free}+m^2]}\right)\Bigg |_{d=3}
&=&\sum_{l=0}^\infty \left(2 l+ 1 \right) \left\{
\ln \left(\frac{\psi_{(l)}(\infty)}{\psi^{\rm free}_{(l)}(\infty)}\right)-\frac{\int_0^\infty dr\, r V(r)}{2(l+\frac{1}{2})}\right\}
\label{3d-result}\\ \nn\\
\ln\left(\frac{\det [{\mathcal M}+m^2]}{\det [{\mathcal M}^{\rm free}+m^2]}\right)\Bigg |_{d=4}&=& \nn\\
&&\hskip-3cm \sum_{l=0}^\infty
\left(l+1\right)^2 \left\{
\ln \left(\frac{\psi_{(l)}(\infty)}{\psi^{\rm free}_{(l)}(\infty)}\right)-\frac{\int_0^\infty dr\, r V(r)}{2(l+1)}+\frac{\int_0^\infty dr\, r^3 V(V+2m^2)}{8(l+1)^3}\right\} \nn\\
&&-\frac{1}{8}\int_0^\infty dr\, r^3\, V(V+2m^2) \left[\ln\left(\frac{\mu r}{2}\right)+\gamma+1\right] \quad .
\label{4d-result}
\eea
Here $\gamma$ is Euler's constant, and $\mu$ is a renormalization scale (defined in the next section), which is essential for physical applications, and which arises naturally in even dimensions. A conventional renormalization choice is to take $\mu=m$ in \eqn{2d-result}--\eqn{4d-result}. In each of \eqn{2d-result}--\eqn{4d-result}, the sum over $l$ is convergent once
the indicated subtractions are made. The function $\psi_{(l)}(r)$ is the solution to the radial equation
\bea
\left[{\mathcal M}_{(l)}+m^2\right] \psi_{(l)} (r) &=&0\nn\\
\psi_{(l)}(r)&\sim& r^{l+(d-1)/2} \quad,\quad  {\rm as}\quad r\to
0\quad . \label{psi-l} \eea The function $\psi_{(l)}^{\rm
free}(r)$ is defined similarly, with the same behavior as $r\to
0$, in terms of the operator $[{\mathcal M}^{\rm free}_{(l)}+m^2
]$. Thus, in $d$ dimensions, $\psi_{(l)}^{\rm free}(r)$ is
expressed as a Bessel function: \bea \psi_{(l)}^{\rm free}
(r)=\left(\frac{2}{m}\right)^{l+d/2-1}
\Gamma\left(l+\frac{d}{2}\right) r^{1/2}\, I_{l+d/2-1}(m r)\quad .
\label{psi-free} \eea

Notice that the results \eqn{2d-result}--\eqn{4d-result} state
once again that the determinant is determined by the boundary
values of solutions of $\left[{\mathcal M}+m^2\right]\psi=0$, with
the only additional information being a finite number of integrals
involving the potential $V(r)$. We also stress the computational
simplicity of \eqn{2d-result}--\eqn{4d-result}, as the initial
value problem \eqn{psi-l} is trivial to implement numerically. The
$d=4$ result \eqn{4d-result} was found previously in
\cite{dunnemin} using radial WKB and an angular momentum cutoff
regularization and renormalization \cite{wkb}. Here we present a
different proof using the zeta function approach, and generalize
to other dimensions. In fact, the $d=2$ and $d=3$ results can also
be derived using the radial WKB method of \cite{dunnemin,wkb}.
Furthermore, in Section \ref{sec-feynman} we show how these
results also agree with the Feynman diagrammatic approach, by
showing that the $d=4$ zeta function expression \eqn{4d-result}
agrees precisely with a superficially different, and more
complicated, $d=4$ expression found in \cite{baacke}. Finally, we
note that the results in \eqn{2d-result}--\eqn{4d-result} are for
a {\it generic} radial potential $V(r)$. There are important
physical applications where the potential $V(r)$ is such that the
operator ${\mathcal M}$ has negative and/or zero modes
\cite{dunnemin}, in which case these expressions are modified
slightly, as in \cite{dunnemin} and as discussed below in the
conclusions.

\section{Zeta Function Formalism}
\label{sec-zeta}

The functional determinant can be defined in terms of a zeta
function \cite{ray,hawking,elizalde} for the operator ${\mathcal
M}$. For dimensional reasons, we define
 \bea \zeta_{[{\mathcal
M}+m^2]/\mu^2}(s)=\mu^{2s} \, \zeta_{[{\mathcal M}+m^2]} (s) =
\mu^{2s} \sum_\lambda (\lambda+m^2)^{-s} \quad , \label{zeta-mu}
\eea where the sum is over the spectrum of ${\mathcal M}$, and
$\mu$ is an arbitrary parameter with dimension of a mass.
Physically, $\mu$ plays the role of a renormalization scale. Then
the logarithm of the determinant is defined as
\cite{ray,hawking,elizalde} \bea
\ln \det \left[{\mathcal M}+m^2\right] &\equiv &-\zeta^\prime_{[{\mathcal M}+m^2]/\mu^2}(0)\nn\\
&=&-\ln (\mu^2) \, \zeta_{[{\mathcal
M}+m^2]}(0)-\zeta^\prime_{[{\mathcal M}+m^2]}(0) \quad .
\label{zeta-det} \eea To compute the determinant {\it ratio}, we
define the zeta function difference \bea \zeta(s)&\equiv &
\zeta_{[{\mathcal M}+m^2]}(s)-\zeta_{[{\mathcal M}^{\rm
free}+m^2]}(s) \quad . \label{zeta} \eea Thus we need to compute
the zeta function and its derivative, each evaluated at $s=0$. In
general, the zeta function at  $s=0$ is related to the heat kernel
coefficient, $a_{d/2}({\mathcal P})$, associated with the operator
${\mathcal P}$  \cite{kirstenbook,seel}: \bea
\zeta_{\mathcal P}(0)=a_{d/2}({\mathcal P}) \quad .
\label{heatkernel} \eea For the operator ${\mathcal P} = - \Delta
- E$, these heat kernel coefficients are \cite{kirstenbook} given
in $\reals^d$ by \bea a_1({\mathcal
P})&=&\frac{1}{(4\pi)^{d/2}}\int_{\reals^d} d^dx\, E
\label{h-a1}\\
a_{3/2}({\mathcal P})&=&0
\label{h-a32}\\
a_2({\mathcal P})&=&\frac{1}{(4\pi)^{d/2}}\int_{\reals^d} d^dx\,
\frac{E^2}{2} \quad . \label{h-a2} \eea Thus, setting $E=-V(r) -
m^2$, and $E^{\rm free}=-m^2$, we find \bea \zeta(0) =\cases{
-\frac{1}{2}\int_0^\infty dr\, r V(r) \hskip 3.1cm ,\quad d=2 \cr
0 \hskip 5.75cm   ,\quad d=3\cr \frac{1}{16}\int_0^\infty dr\, r^3
V(V+2 m^2) \hskip 1.75cm ,\quad d=4 \quad .} \label{ren-terms}
\eea

The derivative of the zeta function at $s=0$, $\zeta^\prime(0)$,
can be evaluated using the relation to the Jost functions of
scattering theory \cite{taylor,alfaro};  for the application
of these ideas to the Casimir effect see \cite{mich}. Consider
the radial eigenvalue equation \bea {\mathcal M}_{(l)}\phi_{(l),
p}=p^2 \phi_{(l), p} \quad , \label{eigenfunction} \eea where
${\mathcal M}_{(l)}$ is the Schr\"odinger-like radial operator
defined in \eqn{radial-op}. A distinguished role is played by the so-called
{\it regular solution}, $\phi_{(l),p} (r)$, which
is defined to have the same behavior as $r\to 0$ as the solution
{\it without} potential:
\begin{eqnarray}
\phi_{(l),p} (r) &\sim& \hat j_{l+(d-3)/2} (pr)\quad. \label{besselj}
\\[-10pt]
&{\stackrel {r \to 0}{}}&\nonumber
\end{eqnarray}
Here the spherical Bessel function
$ \hat j_{l+(d-3)/2} $ is defined as
\beq
\hat j_{l+(d-3)/2} (z) = \sqrt{\frac{\pi z} 2 }\, J_{l+d/2-1} (z)\quad .\nn
\eeq
The asymptotic behavior of the regular solution, $\phi_{(l),p}(r)$, as $r\to\infty$ defines the
{\it Jost function}, $f_l(p)$,  \cite{taylor}
\begin{eqnarray}
\phi_{(l),p}(r) &\sim& \frac i 2 \left[ f_l(p)\, \hat h _{l+(d-3)/2}^- (pr) -
        f_l^*(p)\, \hat h _{l+(d-3)/2}^+ (pr) \right]\quad .
        \label{jost}\\[-10pt]
&{\stackrel{r\to\infty}{}}&\nonumber
\end{eqnarray}
Here $\hat h _{l+(d-3)/2}^- (pr)$ and $\hat h _{l+(d-3)/2}^+ (pr)$ are the
Riccati-Hankel functions
 \beq
 \hat h_{ l+(d-3)/2}^+ (z) = i \sqrt{\frac{\pi z} 2 } H_{ l+d/2-1}^{(1)} (z)\quad , \quad
 \hat h_{l+(d-3)/2}^- (z) = -i \sqrt{\frac{\pi z} 2 } H_{l+d/2-1}^{(2)} (z)\quad .\nn
 \eeq

As is well known from scattering theory \cite{taylor,alfaro}, the
analytic properties of the Jost function $f_l (p)$ strongly depend
on the properties of the potential $V(r)$. Analyticity of the Jost
function as a function of $p$ for $\Im p > 0$ is guaranteed, if in
addition to the aforementioned behavior as $r\to\infty$,
we impose $V(r) \sim r^{-2+\ep}$ for $r \to 0$, and continuity of $V(r) $ in $0<r<\infty$ (except perhaps at a finite number of finite discontinuities).
For us, the analytic properties of the Jost function in the upper half plane will be of particular importance because they are related to the shifting of contours in the complex momentum plane.

By standard contour manipulations \cite{kirstenbook}, the zeta function can be expressed in terms of the Jost functions as:
\begin{eqnarray}
\zeta(s)  ={\sin \pi s\over \pi} \sum_{l=0}^{\infty}{\rm deg}(l; d)
\, \int\limits_{m}^{\infty}dk\,\,
[k^2-m^2]^{-s}~\frac{\partial}{\partial k}\ln  f_l (ik)\quad .
\label{zeta-jost}
\end{eqnarray}
This representation is valid for $\Re s > d/2$, and the technical problem is the construction of the analytic continuation of \eqn{zeta-jost} to a neighborhood about $s=0$. If expression \eqn{zeta-jost} were analytic at $s=0$,
then we would deduce that
\begin{eqnarray}
\zeta^\prime(0) = - \sum_{l=0}^{\infty}{\rm deg}(l; d)
\, \ln  f_l (i m) \quad.
\label{zeta-jost-formal}
\end{eqnarray}
From the definition \eqn{jost} of the Jost function, \bea f_l(i
m)=\frac{\phi_{(l),im}(\infty)}{\phi^{\rm free}_{(l),im}(\infty)}
= \frac{\psi_{(l)}(\infty)}{\psi^{\rm free}_{(l)}(\infty)} \quad ,
\label{jost-gelfand} \eea where $\psi_{(l)}(r)$ is defined in
\eqn{psi-l}. Thus, the regulated expression \eqn{zeta-jost-formal}
coincides with the formal partial wave expansion \eqn{formal-sum},
using the Gel'fand-Yaglom result \eqn{gy-result} for each $l$.
However, the expansion \eqn{zeta-jost-formal} is divergent in
positive integer dimensions. In the zeta function approach, the
divergence of the formal sum in \eqn{zeta-jost-formal} is directly
related to the need for analytic continuation of $\zeta(s)$ in $s$
to a region including $s=0$. From \eqn{zeta-jost}, this analytic
continuation relies on the uniform asymptotic behavior of the Jost
function $f_l (i k)$. Denoting this behavior by $f_l^{asym}(i k)$,
the analytic continuation is achieved by adding and subtracting
the leading asymptotic terms of the integrand in \eqn{zeta-jost}
to write
\begin{eqnarray}
\zeta (s) =\zeta_f(s)+\zeta_{as}(s) \quad ,
\label{zeta-split}
\end{eqnarray}
where
\begin{eqnarray}
 \zeta_f (s)&=&\frac{\sin (\pi s)}{\pi}
\sum_{l=0}^{\infty}{\rm deg}(l; d) \, \int\limits_{m}^{\infty}dk\,\, [k^2-m^2]^{-s}
\frac{\partial}{\partial k} [\ln f_l (i k) - \ln f_l ^{asym} (i k) ]\quad ,\quad
\label{zeta-f}
\end{eqnarray}
and
\begin{eqnarray}
 \zeta _{as} (s)
&=&\frac{\sin (\pi s)}{\pi}  \sum_{l=0}^{\infty}{\rm deg}(l; d) \, \int\limits_{m}^{\infty}dk\,\, [k^2-m^2]^{-s}
\frac{\partial}{\partial k} \ln f_l ^{asym} (i k)\quad .
\label{zeta-as}
\end{eqnarray}
Ultimately we are interested in the analytic continuation of $\zeta (s)$ to $s=0$. As many asymptotic terms will be included in $f_l ^{asym}(ik)$ as  are necessary to make $\zeta_f (s)$ as given in (\ref{zeta-f}) analytic around $s=0$. On the other hand, for $\zeta _{as}(s)$ the analytic continuation to $s=0$ can be constructed in closed form using an explicit representation of the asymptotic behavior of the Jost function, derived in the next section.

\subsection{Asymptotics of the Jost Function}
\label{sec-jost}

The asymptotics of the Jost function $f_l (i k)$ follows from standard results in scattering theory \cite{taylor}. The starting point is the integral equation for the regular solution
\beq
\phi_{(l), p} (r) = \hat j_{l+(d-3)/2} (p r) +
 \intl_0^r dr' \,\, {\cal G} _{l, p} (r,r')\, V(r')\, \phi_{(l),p} (r') \quad ,
\label{integral-equation}
\eeq
with the Green's function
\beq {\cal G}_{l, p} (r,r') = \frac i {2p} \left[
    \hat h _{l+(d-3)/2}^- (pr) \hat h _{l+(d-3)/2}^+ (p r') -
    \hat h_{l+(d-3)/2}^+ (p r) \hat h _{l+(d-3)/2}^- (p r') \right]\quad .
    \label{green-l}
\eeq
Asymptotically for $r\to\infty$,
\beq
  \phi_{(l),p} (r) \sim \hat j_{l+(d-3)/2} (pr) +
 \intl_0^\infty dr' \,\, {\cal G}_{l,p}(r,r')\, V(r')\, \phi_{(l),p} (r')\quad .
 \label{integral-asymptotic}
\eeq
Noting that $[\hat h ^\pm _\nu (x)]^* = \hat h ^\mp _\nu (x)$, for $x$ real,
this asymptotic behavior can be written as
\beq
\phi_{(l),p} (r) &\sim & \frac i 2
\left\{ \left[ 1+\frac 1 p \intl_0^\infty dr' \,\,
   \hat h ^+_{l+(d-3)/2} (pr')\, V(r')\, \phi_{(l),p} (r') \right]  \hat h ^-_{l+(d-3)/2}(pr)
           \right. \nn\\
   & &\left.-\left[ 1+\frac 1 p \intl_0^\infty dr' \,\,
   \hat h ^+_{l+(d-3)/2} (pr')\, V(r')\, \phi_{(l),p} (r') \right] ^* \hat h_{l+(d-3)/2}^+ (pr)
          \right\} .\nn
\eeq
Comparing with the definition (\ref{jost}) of the Jost function, we find the following integral equation
for the Jost function:
\beq
f_l (p) =
1+\frac 1 p \intl_0^\infty dr \,\,
          \hat h _{l+(d-3)/2}^+ (pr)\, V(r)\, \phi_{(l), p} (r)   \quad .
          \label{jost-integral}
\eeq
For the zeta function \eqn{zeta-jost} we need the Jost function for imaginary argument, so we rotate using the Bessel
function properties \cite{as}
 \beq I_\nu (z) = e^{-\frac \pi 2 \nu
i} J_\nu (iz) \quad , \quad K_\nu (z) = \frac{\pi i} 2 e^{\frac \pi 2
\nu i} H_\nu ^{(1)} (iz) \quad . \nn
\eeq
Thus \eqn{jost-integral} becomes
\begin{eqnarray}
f_l (ik) =1+\int\limits_0^{\infty} dr\,r\,V(r)\, \phi_{(l),ik} (r)
   K_{l+d/2-1} (kr)\quad .
   \label{jost-imaginary}
\end{eqnarray}
We define a convenient short-hand for the Bessel function index,
\bea
\nu\equiv l+\frac{d}{2}-1\quad ,
\label{nu}
\eea
and write the partial-wave Lippmann-Schwinger integral equation for the regular solution as
\begin{eqnarray}
\phi_{(l),ik} (r) &=& I_{\nu} (kr)
+
\int\limits_0^r dr'\,\, r'\,\,[ I_{\nu } (kr) K_{\nu} (kr')-
 I_{\nu} (kr') K_{\nu} (kr)] V(r') \phi_{(l),ik}(r')\quad . \quad ~
 \label{lippmann}
\end{eqnarray}
This Lippmann-Schwinger equation leads to an iterative expansion for $f_l (ik)$ in powers of the potential $V(r)$.
For dimensions $d\leq 4$, we need at most the  $O(V)$ and $O(V^2)$ terms of $\ln f_l(i k)$:
\begin{eqnarray}
\ln f_l (ik) = \int\limits_0^{\infty}dr\, r\, V(r)
K_{\nu} (kr) I_{\nu} (kr)-\int\limits_0^{\infty}dr\,r\, V(r)
K_{\nu}^2 (kr) \int\limits_0^rdr'\, r' \,
 V(r') I_{\nu}^2 (kr')+ {\cal O} (V^3) \quad . \nn\\
\label{jost-expansion}
\end{eqnarray}
This iterative scheme effectively reduces the calculation of the
asymptotics of the Jost function to the known uniform asymptotics
of the modified Bessel functions $K_\nu$ and $I_\nu$ \cite{as}. To
the required order, for $\nu\to\infty$, $k\to\infty$, with $k/\nu
$ fixed,
\begin{eqnarray}
I_{\nu} (kr   ) K_{\nu} (kr ) &\sim & \frac t {2\nu}
+\frac{t^3}{16\nu^3}\left( 1-6 t^2
+5 t^4 \right) + {\cal O} \left(\frac{1}{\nu^4} \right)\quad ,\nonumber\\
I_{\nu} (k r' ) K_{\nu} (kr)& \sim&
\frac 1 {2\nu} \frac{e^{-\nu (\eta (k) -\eta (kr'/r))}}
{(1+(kr/\nu)^2)^{1/4} (1+(kr'/\nu)^2)^{1/4}}\left[ 1+ {\cal O}  \left(\frac{1}{\nu} \right)\right] \quad ,
\label{uniform}
\end{eqnarray}
where $t\equiv 1/\sqrt{1+(kr/\nu)^2}$, and $\eta (k) \equiv \sqrt{1+(kr/\nu)^2}+\ln [(kr/\nu) /(1$
$+\sqrt{1+(kr/\nu)^2})]$.
The $r'$ integration in the term quadratic in $V$ is performed by
the saddle point method \cite{kirstenbook}. We therefore {\it define} $\ln f_l^{asym}(ik)$ as the $O(V)$ and  $O(V^2)$ parts of this uniform asymptotic expansion:
\begin{eqnarray}
\ln f_l^{asym}(ik)&\equiv &
 \frac 1 {2\nu} \int\limits_0^{\infty}dr\,
\frac{r\,V(r)}{\left[1+\left(\frac {kr}
{\nu}\right)^2 \right]^{1/2}}
\nn\\
&&
+\frac 1 {16\nu^3}\int\limits_0^{\infty}dr\,
\frac{r\,V(r)}{ \left[1+\left(\frac {kr}
{\nu}\right)^2 \right] ^{3/2}}\left[1 -
 \frac 6
  {\left[1+\left(\frac {kr} {\nu}\right)^2 \right] }
+ \frac 5  {\left[1+\left(\frac {kr}
{\nu}\right)^2 \right]^{2}}\right] \nonumber\\
&& -\frac 1 {8\nu^3}\int\limits_0^{\infty}dr\,\,\frac{r^3\, V^2
(r)} { \left[1+\left(\frac {kr} {\nu}\right)^2 \right]^{3/2}} \quad .
\label{asymptotic-jost}
\end{eqnarray}

\subsection{Computing $\zeta_{f}^\prime(0)$}

By construction, $\zeta_f (s)$, defined in \eqn{zeta-f}, is now well
defined at $s=0$, and we find
\begin{eqnarray}
 \zeta_f' (0) &=& -\sum_{l=0}^{\infty}{\rm deg}(l; d)
       \left[\ln f_l (i m) -\ln f_l^{asym} (i m)
\right] \quad .
\label {zetaprime-f}
\end{eqnarray}
This form is suitable for straightforward numerical computation,
as the Jost function $f_l(i m)$ can be computed using \eqn{psi-l}
and \eqn{jost-gelfand}, while $\ln f_l^{asym}(im)$ can be computed
using \eqn{asymptotic-jost}. With the subtraction of $\ln
f_l^{asym}(im)$ in \eqn{zetaprime-f}, the $l$ sum is now
convergent.

However, it is possible to find an even simpler expression. It
turns out that the subtraction in \eqn{zetaprime-f} is an
over-subtraction. To see this, expand $\ln f_l^{asym} (i m)$ into
its large $l$ behavior as follows:
\begin{eqnarray}
\ln f_l^{asym}(i m)&\sim &
 \frac 1 {2\nu} \int\limits_0^{\infty}dr\,r\,V(r)
 -\frac 1 {8\nu^3}\int\limits_0^{\infty}dr\,
r^3\,V\,(V+2m^2)\nn\\
&&+\frac 1 {2\nu} \int\limits_0^{\infty}dr\,
r\,V(r)\left\{\left[1+\left(\frac {m r}
{\nu}\right)^2 \right]^{-1/2}-1+\frac{1}{2}\left(\frac{mr}{\nu}\right)^2\right\}\nn\\
&&+\frac 1 {16\nu^3}\int\limits_0^{\infty}dr\,
\frac{r\,V(r)}{ \left[1+\left(\frac {m r}
{\nu}\right)^2 \right] ^{3/2}}\left[1 -
 \frac 6
  {\left[1+\left(\frac {m r} {\nu}\right)^2 \right] }
+ \frac 5  {\left[1+\left(\frac {m r}
{\nu}\right)^2 \right]^{2}}\right] \nonumber\\
&&-\frac 1 {8\nu^3}\int\limits_0^{\infty}dr\,r^3\, V^2(r)
\left\{\left[1+\left(\frac {m r} {\nu}\right)^2
\right]^{-3/2}-1\right\}. \label{large-l}
\end{eqnarray}
The first term is $O\left(\frac{1}{l}\right)$, and the second is
$O\left(\frac{1}{l^3}\right)$, while the remaining terms are all
$O\left(\frac{1}{l^5}\right)$. In dimensions $d\leq 4$, the
degeneracy factor ${\rm deg}(l; d)$ is at most quadratic in $l$,
and so these last terms are finite when summed over $l$ in
\eqn{zetaprime-f}. (In fact, in $d=2$ and $d=3$, the
$O\left(\frac{1}{l^3}\right)$ terms are also finite when summed
over $l$.) In the next section we show that these finite terms
cancel exactly against corresponding terms arising in the
evaluation of  $\zeta_{as}^\prime(0)$. Thus, for
$\zeta^\prime(0)=\zeta_f^\prime(0)+\zeta^\prime_{as}{(0)}$, we
only actually need to subtract the {\it leading} large $l$ terms
in \eqn{large-l}, rather than the full asymptotics in
\eqn{asymptotic-jost}.

\subsection{Computing $\zeta_{as}^\prime(0)$}

The explicit form of the asymptotic terms in \eqn{asymptotic-jost}
provides the analytic continuation to $s=0$ of $\zeta_{as} (s)$,
as defined in \eqn{zeta-as}. The $k$ integrals are done using
\begin{eqnarray}
{\int\limits_{m}^{\infty}dk\,\, [k^2-m^2]^{-s}
\frac{\partial}{\partial k}}\left[1+\left(\frac {kr}
{\nu}\right)^2 \right]
     ^{-\frac{n}{2}}=  -\frac{\Gamma (s+\frac{n}{2})
\Gamma (1-s)}{\Gamma (n/2)}
 \frac{\left(\frac{\nu}{m r}\right)^n
m^{-2s}}{\left(1+
\left(\frac{\nu}{m r}\right)^2\right)^{s+\frac{n}{2}}}\quad .
\label{k-integral}
\end{eqnarray}
Therefore, we find
\bea
\zeta_{as}(s)&=&-\sum_{l=0}^\infty {\rm deg}(l; d)\left[\int_0^\infty dr\, r^{1+2s}\, V(r)\left\{ \frac{1}{2}\frac{\Gamma(s+\frac{1}{2})}{\Gamma(s)\Gamma(\frac{1}{2})}\frac{\nu^{-1-2s}}{\left(1+\left(\frac{mr}{\nu}\right)^2\right)^{s+1/2}}\right.\right.\nn\\
&&+\left.\left.\frac{1}{16}\frac{\Gamma(s+\frac{3}{2})}{\Gamma(s)\Gamma(\frac{3}{2})}\frac{\nu^{-3-2s}}{\left(1+\left(\frac{mr}{\nu}\right)^2\right)^{s+3/2}}-\frac{3}{8}\frac{\Gamma(s+\frac{5}{2})}{\Gamma(s)\Gamma(\frac{5}{2})}\frac{\nu^{-3-2s}}{\left(1+\left(\frac{mr}{\nu}\right)^2\right)^{s+5/2}}\right.\right.\nn\\
&&+\left.\left. \frac{5}{16}\frac{\Gamma(s+\frac{7}{2})}{\Gamma(s)\Gamma(\frac{7}{2})}\frac{\nu^{-3-2s}}{\left(1+\left(\frac{mr}{\nu}\right)^2\right)^{s+7/2}}\right\}\right.\nn\\
&&-\left. \int_0^\infty dr\, r^{3+2s}\, V^2(r)\,
\frac{1}{8}\frac{\Gamma(s+\frac{3}{2})}{\Gamma(s)\Gamma(\frac{3}{2})}\frac{\nu^{-3-2s}}{\left(1+\left(\frac{mr}{\nu}\right)^2\right)^{s+3/2}}
\right] \quad .
\label{zeta-as-full}
\eea
We now subtract sufficiently many terms inside the $l$ sum to ensure the analytic continuation of $\zeta_{as}(s)$ to $s=0$. The added back terms produce Riemann zeta function terms, such as $\zeta_R(2s+1)$, whose analytic continuation is immediate. For example, in $d=4$, where $\nu=l+1$, and ${\rm deg}(l; 4)=(l+1)^2=\nu^2$,  the first term in \eqn{zeta-as-full} involves the function
\bea
{\mathcal R}_1(s)\equiv
\frac{\Gamma(s+\frac{1}{2})r^{2s}}{\Gamma(s)\Gamma(\frac{1}{2})}\sum_{\nu=1}^\infty
\frac{\nu^{1-2s}}{\left(1+\left(\frac{mr}{\nu}\right)^2\right)^{s+1/2}}
\quad .
\label{r1}
\eea
The analytic continuation of this function to $s=0$ is
\bea
{\mathcal R}_1(s)&=&\frac{\Gamma(s+\frac{1}{2})r^{2s}}{\Gamma(s)\Gamma(\frac{1}{2})}\left[ \sum_{\nu=1}^\infty \nu^{1-2s}\left\{ \left(1+\left(\frac{mr}{\nu}\right)^2\right)^{-s-1/2}-1+\left(s+\frac{1}{2}\right)\left(\frac{mr}{\nu}\right)^2 \right\} \right.\nn\\
&&\left. +\zeta_R(2s-1)-\left(s+\frac{1}{2}\right)(mr)^2 \zeta_R(2s+1)\right]\quad .
\label{r1-contin}
\eea
A straightforward computation yields the derivative at $s=0$:
\bea
{\mathcal R}_1^\prime(0)&=&
\sum_{\nu=1}^\infty \nu\left\{ \left(1+\left(\frac{mr}{\nu}\right)^2\right)^{-1/2}-1+\frac{1}{2}\left(\frac{mr}{\nu}\right)^2\right\}\nn\\
&&-\frac{1}{2}(mr)^2\left[\ln\left(\frac{r}{2}\right)+\gamma+1\right]+\zeta_R(-1) \quad .
\eea
Applying this strategy to the remaining terms in \eqn{zeta-as-full} leads to
\bea
\zeta^\prime_{as}(0)\Bigg |_{d=4}&=& \frac{1}{8}\int_0^\infty dr\, r^3\, V(V+2m^2)\left[\ln\left(\frac{r}{2}\right)+\gamma+1\right]\nn\\
&&-\int_0^\infty dr\,r \, V(r)\left\{ \frac{1}{2}\sum_{\nu=1}^\infty \nu\left[ \left(1+\left(\frac{mr}{\nu}\right)^2\right)^{-1/2}-1+\frac{1}{2}\left(\frac{mr}{\nu}\right)^2\right] \right. \nn\\
&&\left. \hskip -2cm +\frac{1}{16}\sum_{\nu=1}^\infty \frac{1}{\nu}\left[ \left(1+\left(\frac{mr}{\nu}\right)^2\right)^{-3/2}
-6 \left(1+\left(\frac{mr}{\nu}\right)^2\right)^{-5/2} +5 \left(1+\left(\frac{mr}{\nu}\right)^2\right)^{-7/2}\right]
\right\}\nn\\
&&+\frac{1}{8}\int_0^\infty dr\, r^3\, V^2(r) \sum_{\nu=1}^\infty
\frac{1}{\nu}\left[
\left(1+\left(\frac{mr}{\nu}\right)^2\right)^{-3/2}-1\right] \quad
. \label{4d-zetaprime} \eea Notice that the terms involving
summation over $\nu$ cancel exactly against identical terms in
$\zeta^\prime_f(0)$ from  \eqn{large-l}, after those terms are
summed over $l$ with the $d=4$ degeneracy factor $\nu^2=(l+1)^2$.
Furthermore, note that the $\ln r$ term inside the integral on the
first line of \eqn{4d-zetaprime} is precisely of the same form as
the renormalization term in \eqn{ren-terms}, so the $\ln\mu$ in
\eqn{zeta-det} combines with $\ln r$ to form the dimensionless
combination $\ln (\mu r)$ in \eqn{4d-result}.

The analogous computation in $d=3$, with degeneracy factor ${\rm deg}(l; 3)=(2l+1)=2\nu$, leads to
\bea
\zeta^\prime_{as}(0)\Bigg |_{d=3}&=&
-\int_0^\infty dr\,r \, V(r)\left\{ \sum_{l=0}^\infty \left[ \left(1+\left(\frac{mr}{l+\frac{1}{2}}\right)^2\right)^{-\frac{1}{2}}-1\right] \right. \nn\\
&&\left. \hskip -2cm +\frac{1}{8}\sum_{l=0}^\infty \frac{1}{\left(l+\frac{1}{2}\right)^2}\left[ \frac{1}{\left(1+\left(\frac{mr}{l+\frac{1}{2}}\right)^2\right)^{3/2}}
-\frac{6}{\left(1+\left(\frac{mr}{l+\frac{1}{2}}\right)^2\right)^{5/2}} +
\frac{5}{\left(1+\left(\frac{mr}{l+\frac{1}{2}}\right)^2\right)^{7/2}} \right]
\right\}\nn\\
&&+ \frac{1}{4}\int_0^\infty dr\, r^3\, V^2(r) \sum_{l=0}^\infty
\frac{1}{\left(l+\frac{1}{2}\right)^2}
\left(1+\left(\frac{mr}{l+\frac{1}{2}}\right)^2\right)^{-3/2}
\label{3d-zetaprime} \eea In this case, {\it all} terms in
\eqn{3d-zetaprime} cancel against corresponding terms in
\eqn{large-l}, after summing over $l$ with degeneracy factor
$2\nu$ in $d=3$. The only remaining uncancelled term in
\eqn{large-l} is the first term, which is linear in $V$, and is
the subtraction shown in \eqn{3d-result}. This shows that  in
dimension $d=3$ we did not actually need to expand $\ln
f_l^{asym}(i k)$ to $O(V^2)$ in the first place; the $O(V)$ term
would have been sufficient.

The $d=2$ case is slightly different, as we need to separate the
$l=0$ term from the sum. Here $\nu=l$, and the degeneracy factor
is $1$ for $l=0$, and $2$ for $l\geq 1$. Thus, instead of
\eqn{zetaprime-f} we have \bea \zeta '_f(0)\Bigg |_{d=2}=-\ln
f_0(im)-\sum_{l=1}^\infty 2 \left[\ln f_l(im)-\ln
f_l^{asym}(im)\right]\quad . \label{2d-zetaprime-f} \eea And in
two dimensions \eqn{zeta-as} is \bea \zeta_{as}(s)\Bigg |_{d=2} =
\frac{\sin (\pi s)}{\pi}  \sum_{l=1}^{\infty} 2
\int\limits_{m}^{\infty}dk\,\, [k^2-m^2]^{-s}
\frac{\partial}{\partial k} \ln f_l ^{asym} (i k) \quad .
\label{2d-split} \eea Then the analogous computation in $d=2$
leads to \bea
\zeta^\prime_{as}(0)\Bigg |_{d=2}&=& - \int_0^\infty dr\, r\, V(r)\left[\ln\left(\frac{r}{2}\right)+\gamma\right]\nn\\
&&-\int_0^\infty dr\,r \, V(r)\left\{ \sum_{l=1}^\infty \frac{1}{l} \left[ \left(1+\left(\frac{mr}{l}\right)^2\right)^{-1/2}-1\right] \right. \nn\\
&&\left. \hskip -2cm +\frac{1}{8}\sum_{l=1}^\infty \frac{1}{l^3}\left[ \left(1+\left(\frac{mr}{l}\right)^2\right)^{-3/2}
-6 \left(1+\left(\frac{mr}{l}\right)^2\right)^{-5/2} +5 \left(1+\left(\frac{mr}{l}\right)^2\right)^{-7/2}\right]
\right\}\nn\\
&&+\frac{1}{4}\int_0^\infty dr\, r^3\, V^2(r) \sum_{l=1}^\infty
\frac{1}{l^3}\left(1+\left(\frac{mr}{l}\right)^2\right)^{-3/2}\quad
. \label{2d-zetaprime} \eea As in the $d=4$ case, all terms
involving $l$ summation cancel exactly against identical terms in
\eqn{large-l}, after summing those over $l$, with the $d=2$
degeneracy factors. As in $d=3$, the only remaining uncancelled
term in  \eqn{large-l} is the first term, which is linear in $V$,
and is the subtraction shown in \eqn{2d-result}. This shows that
also  in dimension $d=2$, we did not need to expand $\ln
f_l^{asym}(i k)$ to $O(V^2)$ in the first place; the $O(V)$ term
would have been sufficient.

\section{Comparison With Feynman Diagram Approach}
\label{sec-feynman}

In this section we show that our zeta function computation is equivalent to the Feynman diagrammatic expansion for the logarithm of the determinant \cite{salam,baacke}, although the zeta function approach provides a much simpler form of the final expression. Consider regulating the determinant with dimensional regularization. The perturbative expansion in powers of the potential $V$ is \cite{salam}
\bea
\ln\left(\frac{\det \left[{\mathcal M}+m^2\right]}{\det \left[{\mathcal M}^{\rm free}+m^2\right]}\right) & \equiv& \sum_{k=1}^\infty \frac{(-1)^{k+1}}{k}\, A^{(k)}\nn\\
&=&\quad
\firstorder\quad -\frac{1}{2} \quad
\secondorder \quad +\frac{1}{3} \quad
\thirdorder\quad +\dots\qquad ,
\label{feynman}
\eea
where the dots denote insertions of the potential $V$.
Alternatively, we can expand the dimensionally regulated determinant in partial waves as
\bea
\ln\left(\frac{\det \left[{\mathcal M}+m^2\right]}{\det \left[{\mathcal M}^{\rm free}+m^2\right]}\right) &=& \sum_{l=0}^\infty {\rm deg}(l; d)\, \ln\left(\frac{\det \left[{\mathcal M}_{(l)}+m^2\right]}{\det \left[{\mathcal M}^{\rm free}_{(l)}+m^2\right]}\right) \nn\\
&=& \sum_{l=0}^\infty {\rm deg}(l; d)\,\ln f_l(i m) \quad .
\label{partialwave}
\eea
In the Feynman diagrammatic approach \cite{baacke}, the first two order terms, $A^{(1)}$ and $A^{(2)}$, are separated out:
\bea
\ln\left(\frac{\det \left[{\mathcal M}+m^2\right]}{\det \left[{\mathcal M}^{\rm free}+m^2\right]}\right) &=& \sum_{l=0}^\infty {\rm deg}(l; d)\left\{ \ln f_l(i m) - \left[\ln f_l(i m)\right]_{O(V)} - \left[\ln f_l(i m)\right]_{O(V^2)}\right\}\nn\\
&&+A^{(1)}-\frac{1}{2} A^{(2)} \quad .
\label{feynman-subtractions} \eea With these subtractions, the $l$
sum is now finite, and the divergence lies in the dimensionally
regulated Feynman diagrams $A^{(1)}$ and $A^{(2)}$. We now show
how to relate the expression \eqn{feynman-subtractions} to the
zeta function approach.

Using dimensional regularization, the first order Feynman diagram
is
\bea A^{(1)}&=&\int d^dx\, V(x)\, \lim_{x\to y} G(x, y) \quad .
\label{a1}
\eea
The Helmholtz Green's function in $d$ dimensions
is
\bea G(x, y)=\frac{m^{d-2}}{(2\pi)^{d/2}}\,\frac{K_{d/2-1}\left(m\, |x-y|\right)}{\left(m\, |x-y|\right)^{d/2-1}} \quad .
\label{x-green}
\eea
We now use the
Gegenbauer expansion \cite{watson}
\bea
\frac{K_\nu(|x-y|)}{|x-y|^\nu} = 2^\nu \Gamma(\nu)
\sum_{l=0}^\infty  \left( l+\nu \right) \frac{K_{l+\nu}(r)}{r^\nu}
\frac{I_{l+\nu}(r^\prime)}{(r^\prime)^\nu} C_{l}^{\nu}(\cos
\theta) \quad ,
\label{gegenbauer}
\eea
where $|x-y|=\sqrt{r^2+(r^\prime)^2-2 r\, r^\prime \cos\theta}$. As $x\to y$,  noting that $C_l^{d/2-1}(1)=\left(\matrix{l+d-3\cr
l}\right)$, we find
\bea
A^{(1)}&=&\int_0^\infty dr\,r \, V(r)\sum_{l=0}^\infty \frac{(2l+d-2)}{(d-2)} \left(\matrix{l+d-3\cr l}\right) K_{l+d/2-1}(m r)\, I_{l+d/2-1}(m r)\nn\\
&=& \sum_{l=0}^\infty {\rm deg}(l; d)\left[\ln f_l(i m)\right]_{O(V)}\quad ,
\label{a1-comp}
\eea
in agreement with the $O(V)$ term in the iterative expansion for $\ln f_l(im)$ in \eqn{jost-expansion}.

Similarly, the second order Feynman diagram is
\bea
A^{(2)}&=& \int d^d x \int d^d y \, V(x) G(x, y) V(y) G(y, x) \quad .
\label{a2}
\eea
Using the Gegenbauer expansion \eqn{gegenbauer},
together with the identity \cite{gradsteyn}
\bea
\int_0^\pi d\theta \left(\sin \theta\right)^{ 2\nu}C_l^\nu
(\cos\theta)C_{l^\prime}^\nu (\cos\theta) =\delta_{l\, l^\prime}\,
\frac{\pi 2^{1-2\nu} \Gamma(2\nu+l)}{l!\left(l+\nu\right)
\Gamma^2\left(\nu\right)} \quad , \label{geg-integral}
\eea
we find
\bea
A^{(2)}&=&  2\sum_{l=0}^\infty {\rm deg}(l; d) \int_0^\infty dr\, r\, V(r) K_{l+d/2-1}^2(m r) \int_0^r dr^\prime r^\prime V(r^\prime) I_{l+d/2-1}^2(m r^\prime)\nn\\
&=& -2 \sum_{l=0}^\infty {\rm deg}(l; d)\left[\ln f_l(i
m)\right]_{O(V^2)} \quad. \label{a2-comp} \eea Thus,
$-\frac{1}{2}A^{(2)}$ agrees with the $O(V^2)$ term in the
iterative expansion for $\ln f_l(im)$ in \eqn{jost-expansion},
when summed over $l$ with the appropriate degeneracy factor.
Therefore, the Feynman diagrammatic expression
\eqn{feynman-subtractions} is indeed equivalent to the zeta
function expression \eqn{zeta-jost-formal}, with dimensional
regularization.

Now compare also the finite parts. In the Feynman diagrammatic
approach \cite{baacke}, the finite renormalized logarithm of the
determinant ratio is defined by the subtractions in
\eqn{feynman-subtractions}, together with the finite renormalized
form of the first two Feynman diagrams, $A^{(1)}$ and $A^{(2)}$.
For definiteness we consider the $d=4$ case, in order to compare
with previous work \cite{baacke,dunnemin}. Then the renormalized
logarithm of the determinant ratio is \cite{baacke}
\bea
\ln\left(\frac{\det \left[{\mathcal M}+m^2\right]}{\det \left[{\mathcal M}^{\rm free}+m^2\right]}\right)  &=& \sum_{l=0}^\infty  (l+1)^2\left\{ \ln f_l(i m) - \left[\ln f_l(i m)\right]_{O(V)} - \left[\ln f_l(i m)\right]_{O(V^2)}\right\}\nn\\
&&+A^{(1)}_{\rm fin}-\frac{1}{2} A^{(2)}_{\rm fin} \nn\\
&=& \sum_{l=0}^\infty  (l+1)^2\left\{ \ln \left(\frac{\psi_{(l)}(\infty)}{\psi_{(l)}^{\rm free}(\infty)}\right) - \left[\ln f_l(i m)\right]_{O(V)} - \left[\ln f_l(i m)\right]_{O(V^2)}\right\}\nn\\
&&+A^{(1)}_{\rm fin}-\frac{1}{2} A^{(2)}_{\rm fin} \quad ,
\label{baacke-split}
\eea
where we have used \eqn{jost-gelfand} to identify $\ln f_l(i m)$ with $\psi_{(l)}(\infty)/\psi^{\rm free}_{(l)}(\infty)$.
In \cite{baacke}, the subtraction terms are defined as
\bea
\left[\ln f_l(i m)\right]_{O(V)} &\equiv &h_l^{(1)}(\infty)\nonumber\\
\left[\ln f_l(i m)\right]_{O(V^2)}&\equiv &h^{(2)}_l(\infty)
-\frac{1}{2}\left(h^{(1)}_l(\infty)\right)^2 \quad ,
\label{h-subtractions}
\end{eqnarray}
where $h_l^{(k)}(r)$ is the solution to the differential equation
\bea
\left[\frac{d^2}{d r^2}+\left(\frac{2\,m\, I_{l+1}^\prime(m r)}{I_{l+1}(m r)}+\frac{1}{r}\right)\frac{d}{dr}\right]\, h_l^{(k)}(r) &=&V(r)\, h_l^{(k-1)}(r) \nn\\
h_l^{(k)}(0)=0 \quad &,& h_l^{(k)\prime}(0)=0\nn\\
h_l^{(0)}&\equiv& 1 \quad .
\label{h-equations}
\eea
Comparing with \eqn{jost-expansion}, we can alternatively express these
subtracted terms as
\begin{eqnarray}
\left[\ln f_l(i m)\right]_{O(V)} &=& \int\limits_0^{\infty}dr\, r\, V(r)
K_{\nu} (m r) I_{\nu} (kr)\nonumber\\
\left[\ln f_l(i m)\right]_{O(V^2)}&=&-\int\limits_0^{\infty}dr\, r\, V(r) K_{\nu}^2 (m r)
\int\limits_0^rdr'\, r'\,
 V(r') I_{\nu}^2 (m r') \quad .
\label{integral-subtractions}
\end{eqnarray}
Finally, the finite contributions in \eqn{baacke-split} from the first and second order Feynman diagrams in the $\overline{MS}$ scheme are \cite{baacke} (note the small typo in equation (4.32) of \cite{baacke}):
\bea
A_{\rm fin}^{(1)}&=&-\frac{m^2}{8}\int_0^\infty dr\, r^3\, V(r)\nn\\
A_{\rm fin}^{(2)}&=&\frac{1}{128\pi^4}\int_0^\infty dq\, q^3\, \left| \tilde{V}(q)\right|^2\left[2-\frac{\sqrt{4m^2+q^2}}{q}\, \ln\left(\frac{\sqrt{4m^2+q^2}+q}{\sqrt{4m^2+q^2}-q}\right)\right] \quad ,
\label{a-finite}
\eea
where $\tilde{V}(q)$ is the four dimensional Fourier transform of the radial potential $V(r)$.

With the terms subtracted in \eqn{baacke-split} [evaluated using
either \eqn{h-equations} or \eqn{integral-subtractions}], the $l$
sum is  convergent. So this expression yields a finite answer for
the logarithm of the determinant. On the other hand, the actual
subtraction terms and counterterms in \eqn{baacke-split} are
different from those in \eqn{4d-result}, even though the final net
answer for the finite renormalized determinant is numerically the
same. The difference between the two approaches is that in \eqn{baacke-split}
one subtracts the {\it full} $O(V)$ and $O(V^2)$ dependence of $\ln f_l(i m)$,
given in \eqn{jost-expansion}, and then compensates this subtraction with the Feynman
diagram counter-terms whose finite part in the $\overline{MS}$ scheme are given in \eqn{a-finite}.
On the other hand, in the zeta function computation, we subtract just the {\it asymptotic form}
of these first two Feynman diagrams, as in \eqn{asymptotic-jost},
as is required to analytically continue the zeta function to $s=0$. Subsequently,
in the zeta function approach we notice that even this is an over-subtraction,
as part of this asymptotic behavior cancels against
$\zeta_{as}^\prime(0)$, leaving the final expression
\eqn{4d-result}. The regularized form of these zeta function
subtractions is different from the regularized form of the Feynman
diagrammatic subtractions, but the associated counter-terms are
also different, in such a way that the net result for the
determinant is identical. This follows analytically from
\eqn{a1-comp} and \eqn{a2-comp}, and can easily be  confirmed
numerically. This also serves as an explanation of similar effects
noted in one-loop static energy computations \cite{jaffe,noah}.
However, we note that the zeta function expression \eqn{4d-result}
has a significantly simpler form, with the subtractions only
requiring simple integrals involving $V(r)$, while the
subtractions in \eqn{baacke-split} require the more complicated
integrals \eqn{integral-subtractions} [or, equivalently, solving
the differential equations in \eqn{h-equations}], and also require
the Fourier transform of the potential in \eqn{a-finite}.

\section{Conclusions}

To conclude, we have derived simple new expressions,
\eqn{2d-result}--\eqn{4d-result}, for the determinant of a
radially separable partial differential operator of the form
$-\Delta+m^2+V(r)$, generalizing the Gel'fand-Yaglom result
\eqn{gy-result} to higher dimensions. This greatly increases the
class of differential operators for which the determinant can be
computed simply and efficiently. Our derivation uses the zeta
function definition of the determinant, but the same expressions
can be found using the radial WKB approach of \cite{dunnemin,wkb}.
Furthermore, we have shown how these expressions relate to the
Feynman diagrammatic definition of the determinant based on
dimensional regularization \cite{baacke}. These superficially
different expressions are in fact equal, although the zeta
function expression is considerably simpler.

A number of generalizations could be made. First, in certain
quantum field theory applications the determinant may have zero
modes, and correspondingly one is actually interested in computing
the determinant with these zero modes removed. Our method provides
a simple way to compute such determinants. For example, in the
false vacuum decay problem \cite{langer,kobzarev,stone,coleman-fv}
arising in a self-interacting scalar field theory in
$d$-dimensional space-time, the prefactor for the semiclassical
decay rate involves the functional determinant of the fluctuation
operator for quantum fluctuations about a radial classical bounce
solution $\Phi_{cl}(r)$. This fluctuation operator is a radially
separable operator of the form considered in this paper, but there
is a $d$-fold degenerate zero mode in the $l=1$ sector, associated
with translational invariance of the classical bounce solution. It
is straightforward to generalize the analysis of \cite{dunnemin}
for the $d=4$ case to other dimensions, to find that the net
prefactor contribution from these $l=1$ zero modes (including the
collective coordinate factor \cite{gervais,coleman}) has a simple
expression solely in terms of the asymptotic behavior of the
bounce solution: \bea \left(\frac{S[\Phi_{cl}]}{2\pi}\right)^{d/2}
\left(\frac{\det^\prime  \left[{\mathcal
M}_{(l=1)}+m^2\right]}{\det \left[{\mathcal M}^{\rm
free}_{(l=1)}+m^2\right]}
\right)^{-1/2}=\left[\left(2\pi\right)^{d/2-1}\, \Phi_{\infty}\,
\left | \Phi_{cl}^{\prime\prime}(0)\right |\right]^{d/2} \quad.
\label{zero-factor} \eea Here the constant $\Phi_{\infty}$ is
defined by the normalization of the asymptotic large $r$ behavior
of the bounce:   $\Phi_{cl}(r)\sim \Phi_{\infty}\, r^{1-d/2}\,
K_{d/2-1}(r)$. Another important generalization is to include {\it
directly} the matrix structure that arises from Dirac-like
differential operators and from non-abelian gauge degrees of
freedom. The Feynman diagrammatic approach is well developed for
such separable problems \cite{baacke2}; for example it has been
applied to the fluctuations about the electroweak sphaleron
\cite{carson,baacke3}, and to compute the metastability of the
electroweak vacuum \cite{strumia}. More recently, the angular
momentum cut-off method has been used to compute the full mass
dependence of  the fermion determinant in a four dimensional
Yang-Mills instanton background \cite{wkb}, to compute the fermion
determinant in a background instanton in the two dimensional
chiral Higgs model \cite{burnier}, and to address the fluctuation
problem for false vacuum decay in curved space \cite{dunnewang}. A
unified zeta function analysis should be possible, as there is a
straightforward generalization of the Gel'fand-Yaglom result
\eqn{gy-result} to systems of ordinary differential operators
\cite{kirsten}.

\section*{Acknowledgments} GD thanks the US DOE for support
through grant DE-FG02-92ER40716.  KK acknowledges support by
the Baylor University Summer Sabbatical Program and by the Baylor
University Research Committee.


\begin{thebibliography}{12345}


\bibitem{schwinger}
A.~Salam and P.~T.~Matthews,
``Fredholm Theory of Scattering in a given Time Dependent Field'', Phys. Rev. {\bf 90}, 690 (1953);
J.~Schwinger,
``The Theory of Quantized Fields. VI'', Phys.\ Rev.\ {\bf 94}, 1362 (1954).

\bibitem{salam}
  A.~Salam and J.~A.~Strathdee,
  ``Comment On The Computation Of Effective Potentials,''
  Phys.\ Rev.\ D {\bf 9}, 1129 (1974);
  R.~Jackiw,
  ``Functional Evaluation Of The Effective Potential,''
  Phys.\ Rev.\ D {\bf 9}, 1686 (1974);
  J.~Iliopoulos, C.~Itzykson and A.~Martin,
  ``Functional Methods And Perturbation Theory,''
  Rev.\ Mod.\ Phys.\  {\bf 47}, 165 (1975).

  \bibitem{coleman}
S.~R.~Coleman,
``The Uses Of Instantons,''
{\it Lectures delivered at 1977 International School of Subnuclear Physics, Erice:
The Whys of Subnuclear Physics}, Edited by A.  Zichichi, (Plenum Press, 1979).

\bibitem{ray}
D.~B.~Ray and I.~M.~Singer,
``R-Torsion and the Laplacian on Riemannian Manifolds'',
Adv. Math. {\bf 7}, 145 (1971).

\bibitem{hawking}
S.~W.~Hawking, ``Zeta Function Regularization of Path Integrals in Curved Space-time'',
Commun.\ Math.\ Phys.\ {\bf 55}, 133, (1977).

\bibitem{stuart1}
J.~S.~Dowker, ``Functional determinants on spheres and sectors'',
J.\ Math.\ Phys.\ {\bf 35}, 4989 (1994).

\bibitem{elizalde}
  E.~Elizalde, S.~D.~Odintsov, A.~Romeo, A.~A.~Bytsenko and S.~Zerbini,
  {\it Zeta regularization techniques with applications}, (World Scientific, Singapore, 1994).

\bibitem{dhoker}
 E.~D'Hoker and D.~H.~Phong,
  ``On Determinants Of Laplacians On Riemann Surfaces,''
  Commun.\ Math.\ Phys.\  {\bf 104}, 537 (1986).

\bibitem{sarnak}
  P.~Sarnak,
  ``Determinants of Laplacians'',
  Commun.\ Math.\ Phys.\ {\bf 110}, 113 (1987).

\bibitem{kirstenbook}
K. Kirsten, {\it Spectral Functions in Mathematics and Physics}, (Chapman-Hall, Boca Raton, 2002).

\bibitem{gy}
I.~M.~Gelfand and A.~M.~Yaglom,
  ``Integration In Functional Spaces And It Applications In Quantum Physics,''
  J.\ Math.\ Phys.\  {\bf 1}, 48 (1960).

\bibitem{levit}
S. Levit and U. Smilansky, ``A theorem on infinite products of eigenvalues of Sturm-Liouville type operators'', Proc. Am. Math. Soc. {\bf 65}, 299 (1977).

\bibitem{forman}
R. Forman, `` Functional determinants and geometry '',
Invent. Math. {\bf 88}, 447 (1987); Erratum, {\it ibid} {\bf 108}, 453 (1992).

\bibitem{kirsten}
  K.~Kirsten and A.~J.~McKane,
  ``Functional determinants by contour integration methods,''
  Annals Phys.\  {\bf 308}, 502 (2003)
  [arXiv:math-ph/0305010];
 ``Functional determinants for general Sturm-Liouville problems,''
 J.\ Phys.\ A {\bf 37}, 4649 (2004)
  [arXiv:math-ph/0403050].

\bibitem{kleinert}
  H.~Kleinert,
  ``Path Integrals in Quantum Mechanics,  Statistics, Polymer Physics,
  and Financial Markets,'' (World Scientific, Singapore, 2004).

\bibitem{dunnemin}
 G.~V.~Dunne and H.~Min,
  ``Beyond the thin-wall approximation: Precise numerical computation of prefactors in false vacuum decay,''
  Phys.\ Rev.\ D {\bf 72}, 125004 (2005)
  [arXiv:hep-th/0511156].

\bibitem{wkb}
  G.~V.~Dunne, J.~Hur, C.~Lee and H.~Min,
   ``Instanton determinant with arbitrary quark mass: WKB phase-shift method and
  derivative expansion,''
  Phys.\ Lett.\ B {\bf 600}, 302 (2004)
  [arXiv:hep-th/0407222];
  ``Precise quark mass dependence of instanton determinant,''
  Phys.\ Rev.\ Lett.\  {\bf 94}, 072001 (2005)
  [arXiv:hep-th/0410190];
``Calculation of QCD instanton determinant with arbitrary mass,''
  Phys.\ Rev.\ D {\bf 71}, 085019 (2005)
  [arXiv:hep-th/0502087].

\bibitem{baacke}
  J.~Baacke and G.~Lavrelashvili,
  ``One-loop corrections to the metastable vacuum decay,''
  Phys.\ Rev.\ D {\bf 69}, 025009 (2004)
  [arXiv:hep-th/0307202].

 \bibitem{gradsteyn}
I.~S.~Gradshteyn and I.~M.~Ryzhik,
{\it Table of Integrals, Series and Products},
(Academic Press, San Diego, 1980).

\bibitem{seel}
R.~T.~Seeley, ``Complex powers of an elliptic operator'', Singular
Integrals, Chicago 1966. Proc. Sympos. Pure Math. {\bf 10}, 288
(1968), American Mathematics Society, Providence, RI.

\bibitem{taylor}
J.~R.~Taylor,
{\it Scattering Theory}, (Wiley, New York, 1972).

\bibitem{alfaro}
V.~de Alfaro and T.~Regge,
{\it Potential Scattering}, (Wiley, New York, 1965).

\bibitem{mich}
M.~Bordag and K.~Kirsten, ``Vacuum energy in a spherically
symmetric background field'', Phys.\ Rev.\ D {\bf 53}, 5753
(1996).

\bibitem{as}
M.~Abramowitz and I.~Stegun, {\it Handbook of Mathematical
Functions}, (Dover, New York, 1965).

\bibitem{watson}
G.~N.~Watson, {\it Theory of Bessel Functions}, (Cambridge
University Press, 1962), Chapter XI.

\bibitem{jaffe}
  E.~Farhi, N.~Graham, R.~L.~Jaffe and H.~Weigel,
  ``Searching for quantum solitons in a 3+1 dimensional chiral Yukawa  model,''
  Nucl.\ Phys.\ B {\bf 630}, 241 (2002)
  [arXiv:hep-th/0112217].

\bibitem{noah}
   N.~Graham and K.~D.~Olum,
  ``Negative energy densities in quantum field theory with a background
  potential,''
  Phys.\ Rev.\ D {\bf 67}, 085014 (2003)
  [Erratum-ibid.\ D {\bf 69}, 109901 (2004)]
  [arXiv:hep-th/0211244].

  \bibitem{langer}
  J.~S.~Langer,
  ``Theory Of The Condensation Point,''
  Annals Phys.\  {\bf 41}, 108 (1967).

\bibitem{kobzarev}
  M.~B.~Voloshin, I.~Y.~Kobzarev, and L.~B.~Okun,
  ``Bubbles In Metastable Vacuum,''
  Sov.\ J.\ Nucl.\ Phys.\  {\bf 20}, 644 (1975)
  [Yad.\ Fiz.\  {\bf 20}, 1229 (1974)].

  \bibitem{stone}
  M.~Stone,
   ``The Lifetime And Decay Of 'Excited Vacuum' States Of A Field Theory
  Associated With Nonabsolute Minima Of Its Effective Potential,''
  Phys.\ Rev.\ D {\bf 14}, 3568 (1976);
 ``Semiclassical Methods For Unstable States,''
  Phys.\ Lett.\ B {\bf 67}, 186 (1977).


\bibitem{coleman-fv}
  S.~R.~Coleman,
 ``The Fate Of The False Vacuum. 1. Semiclassical Theory,''
  Phys.\ Rev.\ D {\bf 15}, 2929 (1977)
  [Erratum-ibid.\ D {\bf 16}, 1248 (1977)];
  C.~G.~Callan and S.~R.~Coleman,
  ``The Fate Of The False Vacuum. 2. First Quantum Corrections,''
  Phys.\ Rev.\ D {\bf 16}, 1762 (1977).

  \bibitem{gervais}
 J.~L.~Gervais and B.~Sakita,
 ``WKB Wave Function For Systems With Many Degrees Of Freedom: A Unified  View
  Of Solitons And Pseudoparticles,''
  Phys.\ Rev.\ D {\bf 16}, 3507 (1977).

\bibitem{baacke2}
  J.~Baacke,
   ``Numerical Evaluation Of The One Loop Effective Action In Static Backgrounds
   With Spherical Symmetry,''
  %
  Z.\ Phys.\ C {\bf 47}, 263 (1990);
 ``The effective action of a spin 1/2 field in the background of a chiral soliton'',   Z. Phys. C {\bf 53}, 407 (1992).

 \bibitem{carson}
  L.~Carson, X.~Li, L.~D.~McLerran and R.~T.~Wang,
   ``Exact Computation Of The Small Fluctuation Determinant Around A Sphaleron,''
  %
  Phys.\ Rev.\ D {\bf 42}, 2127 (1990).

  \bibitem{baacke3}
  J.~Baacke and S.~Junker,
   ``Quantum fluctuations around the electroweak sphaleron,''
  Phys.\ Rev.\ D {\bf 49}, 2055 (1994)
  [arXiv:hep-ph/9308310];
   ``Quantum fluctuations of the electroweak sphaleron: Erratum and addendum,''
  Phys.\ Rev.\ D {\bf 50}, 4227 (1994)
  [arXiv:hep-th/9402078].

\bibitem{strumia}
  G.~Isidori, G.~Ridolfi and A.~Strumia,
   ``On the metastability of the standard model vacuum,''
  Nucl.\ Phys.\ B {\bf 609}, 387 (2001)
  [arXiv:hep-ph/0104016].

  \bibitem{burnier}
  Y.~Burnier and M.~Shaposhnikov,
  ``One-loop fermionic corrections to the instanton transition in two dimensional chiral Higgs model,''
  Phys.\ Rev.\ D {\bf 72}, 065011 (2005)
  [arXiv:hep-ph/0507130].

  \bibitem{dunnewang}
  G.~V.~Dunne and Q.~h.~Wang,
   ``Fluctuations about cosmological instantons,''
  %
  [arXiv:hep-th/0605176], Phys. Rev. D, in press.


  \end{thebibliography}
\end{document}